\documentclass[%
 superscriptaddress,
 aip,
 amsmath,amssymb,
reprint,%
 author-year,%
longbibliography
]{revtex4-2}

\usepackage{
amsmath,
amssymb,
bm,
color,
comment,
dcolumn,
enumerate,
float,
graphicx,
hyperref,
mathrsfs,
tabularx,
titlesec,
soul,
subfigure
}

\definecolor{linkColor}{rgb}{0.7,0,0}
\definecolor{darkred}{rgb}{0.7,0,0}
\hypersetup{pdfborder={0 0 0},colorlinks=true,urlcolor=linkColor,citecolor=linkColor}

\titleformat{\section}
{\color{darkred}\sffamily\bfseries}
{\color{darkred}\thesection}{1em}{}
\titleformat{\subsection}
{\color{darkred}\sffamily\itshape}
{\color{darkred}\thesection}{1em}{}
\titlespacing*{\section}
{0pt}{8pt}{0pt}
\titlespacing*{\subsection}
{0pt}{8pt}{0pt}

\emergencystretch=\maxdimen
\begin{document}

% \title[PRISM 2.0]{An Even Faster Algorithm for Scanning Transmission Electron Microscopy (STEM) Imaging and Diffraction Simulations}

\title[STEM Corrector]{Design of Electrostatic Aberration Correctors for Scanning Transmission Electron Microscopy}

\author{Stephanie M. Ribet}
\email{sribet@lbl.gov}
\affiliation{Department of Materials Science and Engineering, Northwestern University, Evanston, IL, USA}
\affiliation{International Institute of Nanotechnology, Northwestern University, Evanston, IL, USA}
\affiliation{National Center for Electron Microscopy, Molecular Foundry, Lawrence Berkeley National Laboratory, Berkeley, CA, USA}

\author{Steven E. Zeltmann}
\affiliation{Department of Materials Science and Engineering, University of California, Berkeley, Berkeley, CA, USA}

\author{Karen C. Bustillo}
\affiliation{National Center for Electron Microscopy, Molecular Foundry, Lawrence Berkeley National Laboratory, Berkeley, CA, USA}

\author{Rohan Dhall}
\affiliation{National Center for Electron Microscopy, Molecular Foundry, Lawrence Berkeley National Laboratory, Berkeley, CA, USA}

\author{Peter Denes}
\affiliation{Molecular Foundry, Lawrence Berkeley National Laboratory, Berkeley, CA, USA}

\author{Andrew M. Minor}
\affiliation{National Center for Electron Microscopy, Molecular Foundry, Lawrence Berkeley National Laboratory, Berkeley, CA, USA}
\affiliation{Department of Materials Science and Engineering, University of California, Berkeley, Berkeley, CA, USA}

\author{Roberto dos Reis}
\affiliation{Department of Materials Science and Engineering, Northwestern University, Evanston, IL, USA}
\affiliation{International Institute of Nanotechnology, Northwestern University, Evanston, IL, USA}
\affiliation{The NU\textit{ANCE} Center, Northwestern University, Evanston, IL, USA}

\author{Vinayak P. Dravid}
\affiliation{Department of Materials Science and Engineering, Northwestern University, Evanston, IL, USA}
\affiliation{International Institute of Nanotechnology, Northwestern University, Evanston, IL, USA}
\affiliation{The NU\textit{ANCE} Center, Northwestern University, Evanston, IL, USA}
\email{v-dravid@northwestern.edu}

\author{Colin Ophus}
\email{cophus@gmail.com}
\affiliation{National Center for Electron Microscopy, Molecular Foundry, Lawrence Berkeley National Laboratory, Berkeley, CA, USA}

\date{\today}
\begin{abstract}
In a scanning transmission electron microscope (STEM), producing a high-resolution image generally requires an electron beam focused to the smallest point possible. However, the magnetic lenses used to focus the beam are unavoidably imperfect, introducing aberrations that limit resolution. Modern STEMs overcome this by using hardware aberration correctors comprised of many multipole elements, but these devices are complex, expensive, and can be difficult to tune. We demonstrate a design for an electrostatic phase plate that can act as an aberration corrector. The corrector is comprised of annular segments, each of which is an independent two-terminal device that can apply a constant or ramped phase shift to a portion of the electron beam. We show the improvement in image resolution using an electrostatic corrector. Engineering criteria impose that much of the beam within the probe-forming aperture be blocked by support bars, leading to large probe tails for the corrected probe that sample the specimen beyond the central lobe. We also show how this device can be used to create other STEM beam profiles such as vortex beams and beams with a high degree of phase diversity, which improve information transfer in ptychographic reconstructions.
\end{abstract}
% \pacs{PACS Numbers}
\maketitle

\section*{Introduction}
\begin{figure*}
  \begin{center}
    \includegraphics[width=\textwidth]{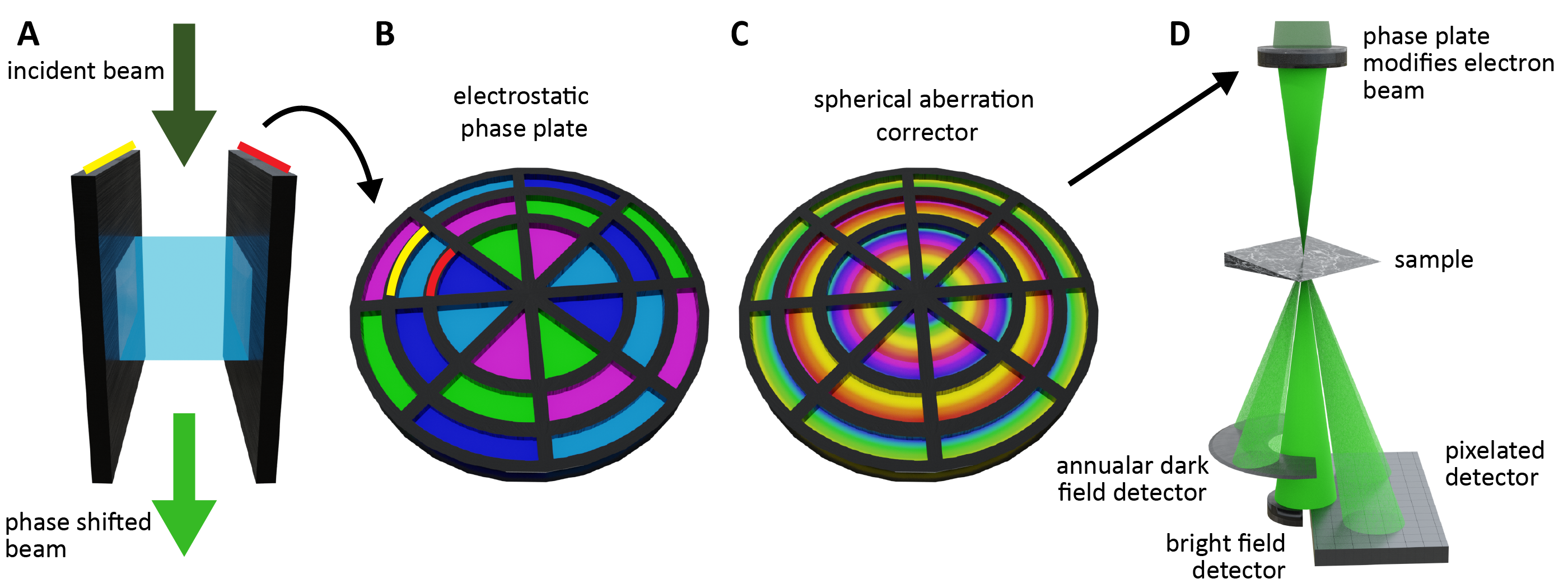}
  \end{center}
  \caption{(A) An electron beam is phase shifted when passed through an electrostatic potential. (B) A series of two-terminal devices can be arranged into a larger phase plate, where the potential in each aperture is tuned independently. Red and yellow lines in (A) and (B) denote terminals. (C) A potential difference between two terminals adds a linear phase ramp, which can be used to correct spherical aberrations. (D) The device is inserted in the probe forming aperture of the STEM to modify the electron beam before interaction with the sample and is compatible with imaging, diffraction, spectroscopy, and 4D-STEM experiments.}
  \label{fig:intro}
\end{figure*} 

Across biological and physical sciences, understanding material systems often requires precise characterization down to the nano or atomic scales. Scanning transmission electron microscopy (STEM) is a key tool to fulfill such requirements, as it has a small probe size that can be used for directly imaging structures and mapping of chemical and physical properties. In STEM, resolution is typically limited by spherical aberrations of the probe-forming lenses. These aberrations are unavoidable and intrinsic to the microscope design -- Scherzer's theorem states that a static, rotationally symmetric magnetic field will always produce spherical aberrations greater than zero. However, Scherzer also pointed out that the resolution can be improved, despite these parasitic aberrations, by balancing defocus and spherical aberrations~\citep{scherzer1936einige}. 

The importance of improving resolution in STEM by limiting aberrations has inspired a series of computational and physical advances that led to the design of multipole aberration correctors~\citep{rose2009history, smith2008development, urban2022progress}.  Nowadays, these correctors have become more widespread, and researchers are routinely able to reach atomic resolution~\citep{batson2002sub, dahmen2009background, urban2022progress}.  Nonetheless, these devices can be expensive and complicated to operate, and there have been continued efforts to find alternative means for aberration correction.~\cite{linck2017aberration} incorporated a diffraction grating into the probe forming aperture of STEM such that the spherical aberrations were canceled in the first-order diffracted beam. Similarly,~\cite{roitman2021shaping} used a thickness-dependent phase shift of silicon nitride thin-films to sculpt a phase plate that acts as an aberration corrector. These simpler devices helped to overcome some of the limitations of conventional aberration correctors, but deploying physical phase plates is still challenging in practice~\citep{malac2021phase}. 

Programmable electrostatic phase plates, as demonstrated in an electron microscope by~\cite{verbeeck2018demonstration}, are an exciting platform for direct control over the electron beam profile. The phase plate has multiple smaller apertures with individual voltage control, so each section has a different phase that can independently impart a shift on the electron beam before interaction with the sample.  This concept allows for remarkable spatial control over the electron beam and can be tuned while inserted in the microscope. This approach to beam sculpting was proposed for aberration correction~\citep{verbeeck2018demonstration}, and further considered by~\cite{vega2023can}. These studies showed that a two terminal device is better than a one terminal device for aberration correction because a linear phase ramp can more readily match the aberration function profile with fewer apertures. In a follow-up work, Vega Ib\'a\~nez, et al.~extended their design to include two terminals with independent voltages, in order to impart phase ramps to the electron beam \citep{vega2023can}, following earlier designs such as \cite{boersch1947kontraste, matsumoto1996phase, schultheiss2006fabrication}.

In this work, we consider the implementation of a programmable phase plate (Fig.~\ref{fig:intro}) and its impact on material characterization in light of realistic experimental constraints.  We highlight the benefits of using such a device both for improving the STEM probe size and the limitations of this approach, mostly arising from the large probe tails due to architectural aperture supports. We illustrate how the programmable phase plate lends itself to the creation of other beam profiles. For example, we show how a programmable phase plate can be used to make a vortex beam or to impart phase diversity in the probe for better ptychographic reconstructions. 

\section*{Theory}
\begin{figure*}
  \begin{center}
    \includegraphics[width=\textwidth]{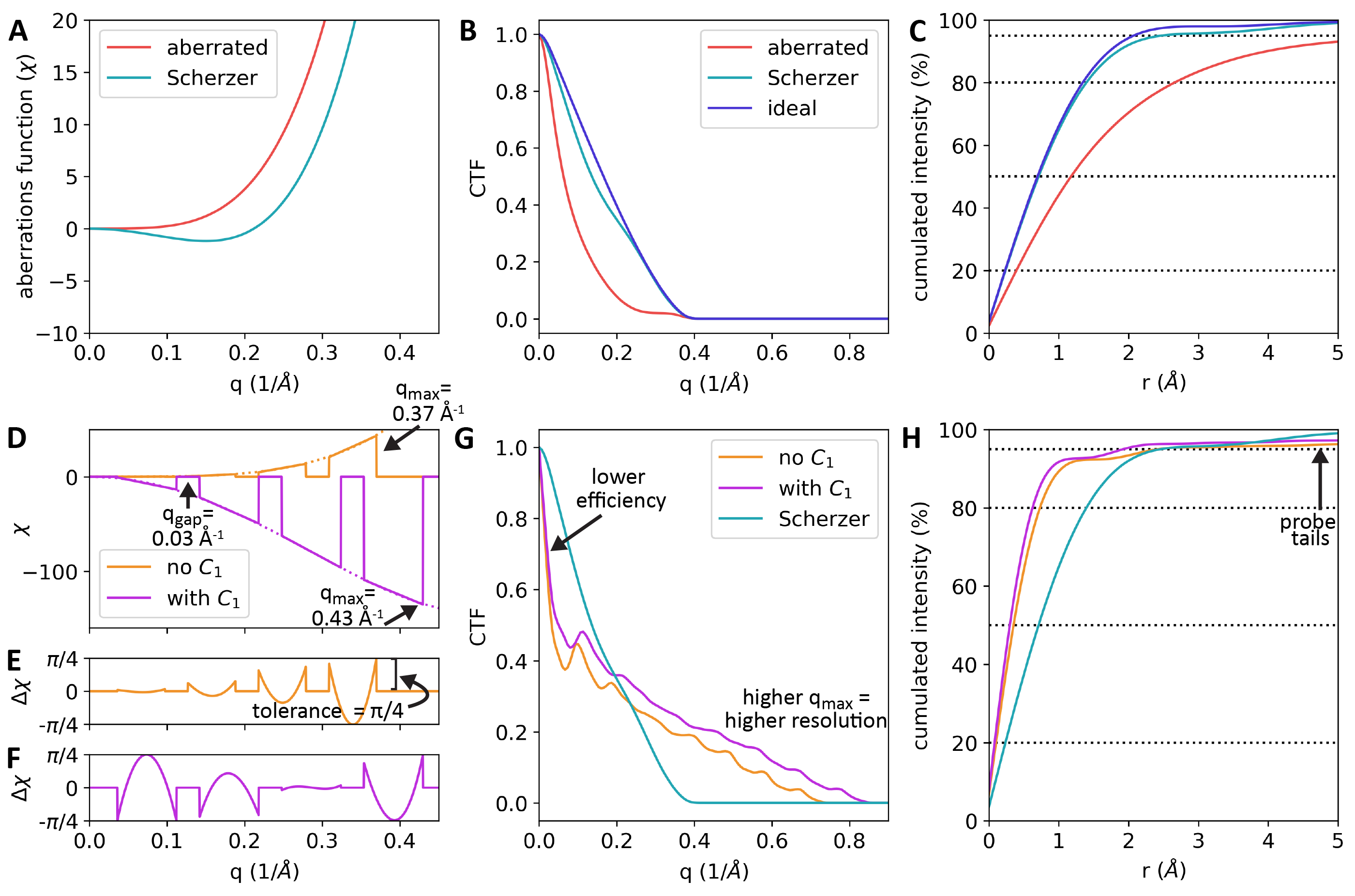}
  \end{center}
  \caption{(A) Aberration functions for STEM probes which contain $C_3$ spherical aberrations either with (Scherzer) and without defocus and (B) resulting CTFs and (C) cumulated probe intensity. $C_1$ can balance the $C_3$ aberrations to make a smaller probe. (D) Aberration functions and corrector profiles and residuals either (E) without and (F) with defocus. Adding $C_1$ helps extend the $q_{max}$ of the device leading to a better (G) CTF and (H) probe profile.}
  \label{fig:scherzer}
\end{figure*}

\begin{figure}
  \begin{center}
    \includegraphics[width=0.5\textwidth]{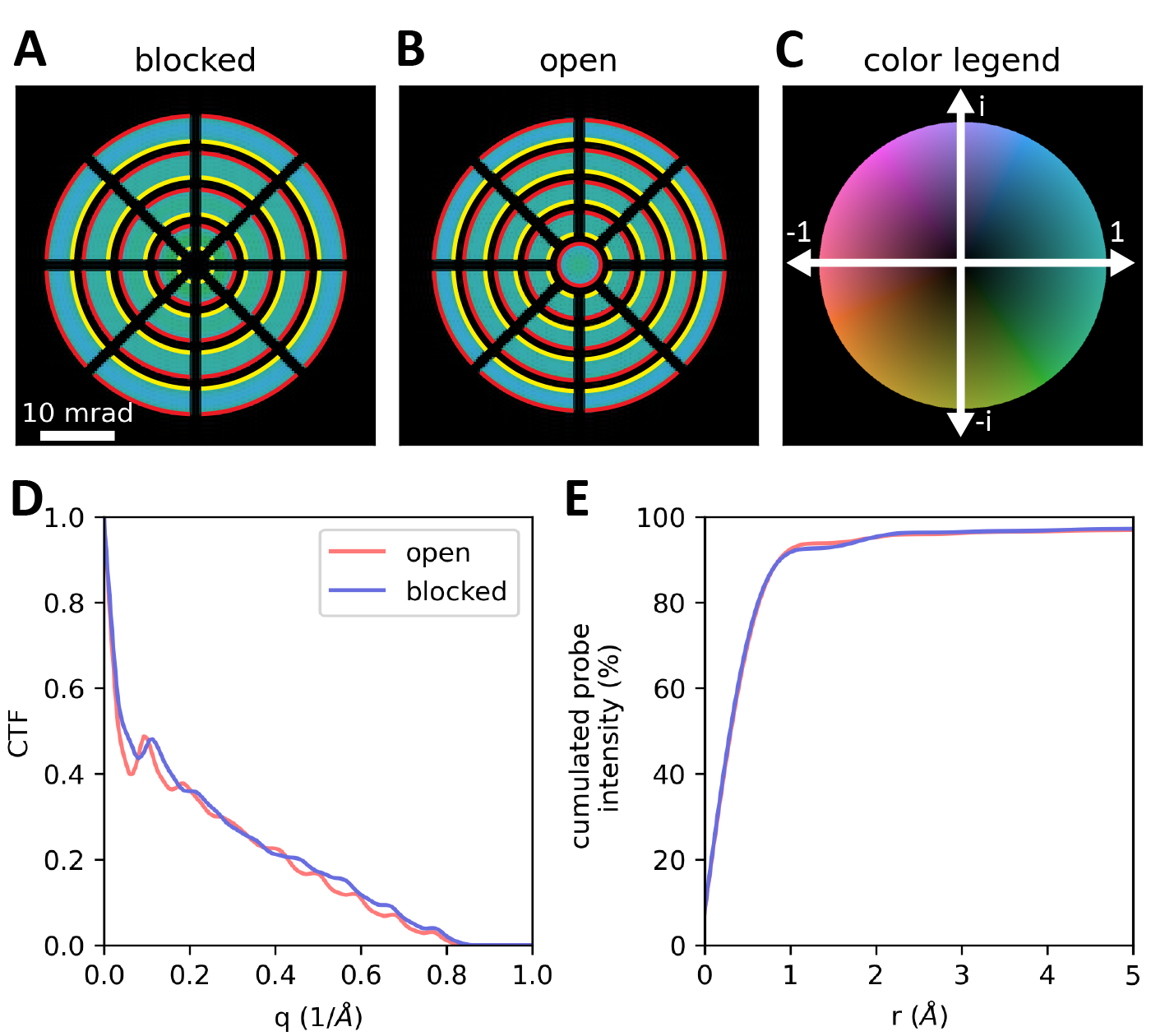}
  \end{center}
  \caption{Two possible designs for the aberration corrector are with a (A) blocked or (B) open center. The (B) open center design accommodates only a one terminal device in the central aperture limiting flexibility. (C) Color legend for (A) and (B) and other probes in the paper. The (D) CTF and (E) probe profiles for the two designs are similar.}
  \label{fig:blocked}
\end{figure}

\begin{figure*}
  \begin{center}
    \includegraphics[width=\textwidth]{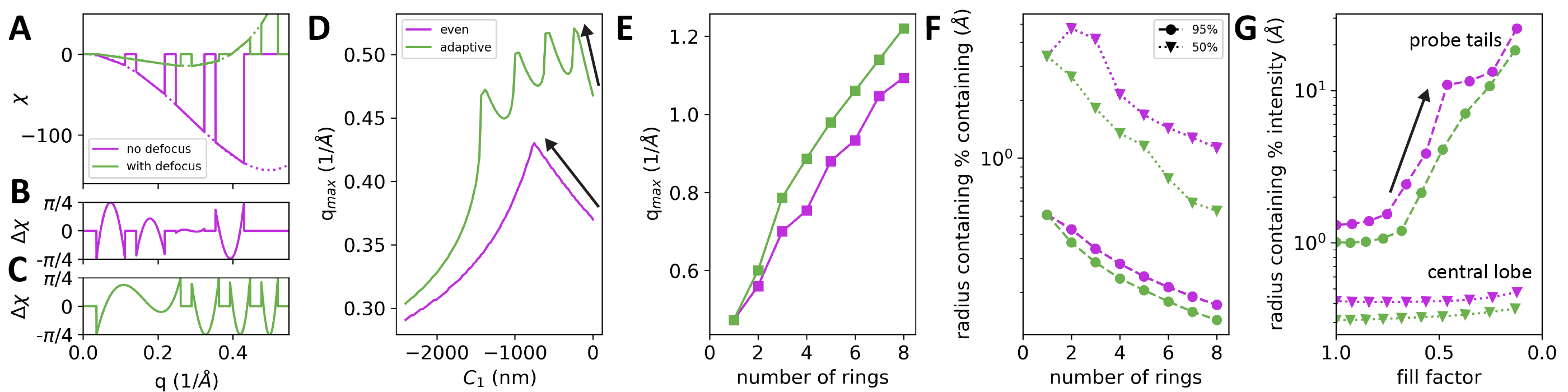}
  \end{center}
  \caption{(A) Aberration function and corrector profile and residuals for (B) even and (C) adaptively spaced design. The adaptively spaced design leads to a higher $q_{max}$. (D) Both can be optimized by starting at 0 defocus and decreasing until the first maximum is reached. (E) As the number of rings is increased, so is the $q_{max}$. (F) As the number of rings is increased the radius containing 50\% of the probe decreased due to the improve resolution for both the evenly and adaptively spaced correctors. The radius containing 95\% of the probe is worse in the evenly spaced corrector. (G) Probe tails are more problematic as the fill factor is increased.}
  \label{fig:adaptive}
\end{figure*}

\begin{figure*}
  \begin{center}
    \includegraphics[width=\textwidth]{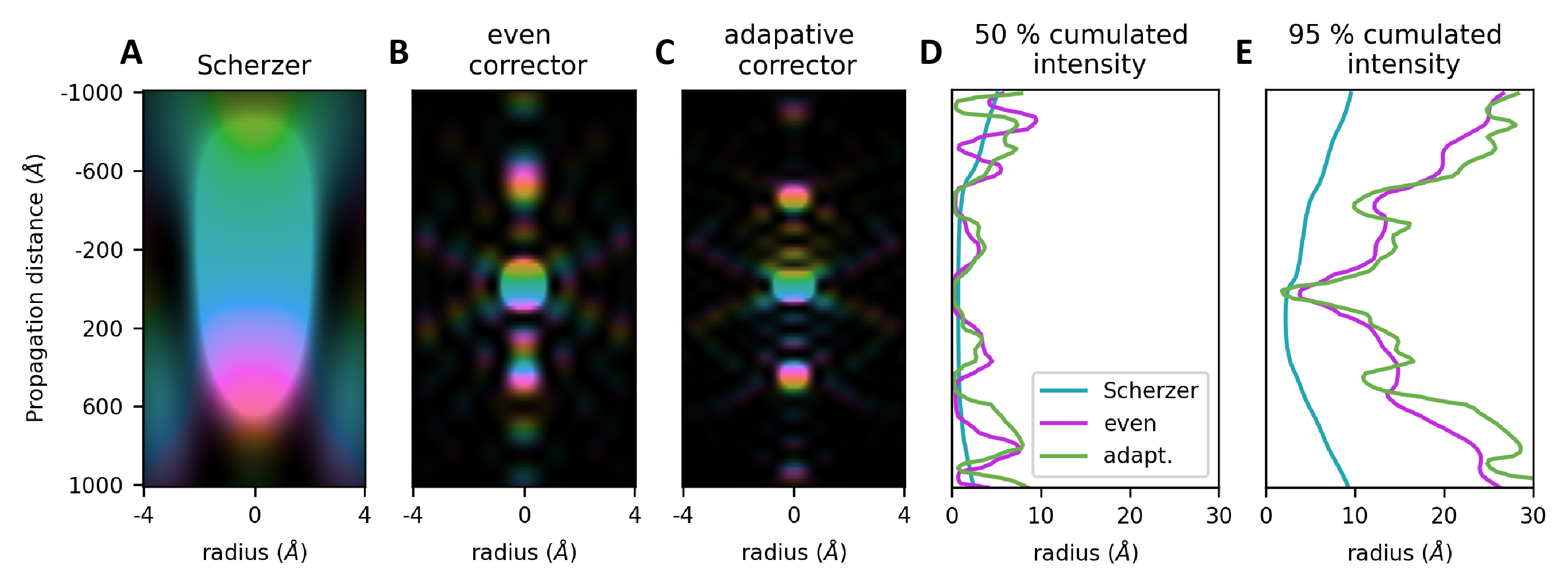}
  \end{center}
  \caption{Probe profiles of the (A) Scherzer condition and corrected probes with the (B) evenly spaced and (C) adaptively spaced phase plate. The corrected probes have a smaller central lobe but modes above and below the plane of the sample that also sample the specimen. (D) 50\% and (E) 95\% probe profiles show long tails for the corrected probes.}
  \label{fig:probe_profile}
\end{figure*} 

Here we optimize the programmable phase plate for aberration correction.
Including only spherical aberrations and defocus, the aberration function can be written as

\begin{equation}
    \chi(q) = \frac{\pi}{2}C_3\lambda^3q^4+\pi C_1 \lambda q^2, 
\end{equation} 

where  $\lambda$ is the de-Broglie wavelength of the electron, $C_3$ is the 3rd order spherical aberration coefficient, and $q$ is the spatial frequency. $C_1 = - \Delta f $, where $\Delta f$ is the defocus. Examples of aberration functions for a microscope with and without defocus to balance the spherical aberrations are shown in Fig.~\ref{fig:scherzer}A. In STEM, the optimal conditions deviate slightly from the Scherzer conditions in TEM, as it is important to balance probe size with probe tails.~\cite{kirkland-parameters} defines the best probe for STEM with the following equations:
\begin{equation}
    C_1 = - 0.87 (C_3\lambda)^{1/2}
    \label{eq:c1}
\end{equation} 
\begin{equation}
    q_{max} = 1.34 (C_3 \lambda^3)^{-1/4}.
\end{equation} 

In Fig.~\ref{fig:scherzer}A-C, we are plotting simulations at 60kV with a maximum scattering angle of 0.21\AA$^{-1}$. More details about simulations throughout the study can be found in the methods section. We use simulations to study the changes to the profile of the probe with aberrations, both with and without a corrector.  In reciprocal space, we can evaluate the efficacy of the transfer of information for various spatial frequencies by computing the contrast transfer function (CTF). The CTF is the Fourier transform of the normalized magnitude squared of the complex real space probe~\citep{goodman2005introduction}.  The CTF of an unaberrated lens has a value of 1 at zero spatial frequency and linearly decreased to zero at the spatial frequency corresponding to twice the semi-convergence angle of the probe (Fig.~\ref{fig:scherzer}B). We observe that by balancing our spherical aberrations with defocus, using the Scherzer condition in Eq.~\ref{eq:c1}, the contrast transfer of information approaches the ideal profile. The maximum $q$ or spatial frequency of information transfer corresponds to when the curve reaches CTF=0, and the deviations from the ideal profile show a loss of efficiency.

There are also many heuristics for evaluating real space probe size that capture not only the radius of the central lobe but also the probe tails, which contribute parasitically to image contrast and resolution. Metrics that highlight information about the central lobe include the probe full width at half maximum and the radius containing 50\% of current~\citep{kirkland-parameters}.  Amplitude plates are known to create longer probe tails, so previous work describing bullseye-patterned apertures defined the STEM probe size as the radius containing 80\% of the total probe intensity~\citep{zeltmann2020patterned}. Finally,~\cite{schnitzer2020optimal} used the Strehl ratio, the ratio of peak intensity between an aberrated and non-aberrated probe, to account for probe size. The real space probe profiles of the same probes are shown in Fig.~\ref{fig:scherzer}C. Again, we observe how balancing the spherical aberrations with defocus can improve the probe size.

Our design for a corrector is based on a two-terminal device, as shown in Fig.~\ref{fig:intro}A, which will be used to create a local electric field across an aperture. A group of these apertures can be patterned into an electrostatic phase plate (Fig.~\ref{fig:intro}B). Further, these apertures will have a linear phase profile, which can be used to approximate segments of the spherical aberration function to correct the probe (Fig.~\ref{fig:intro}C). This phase plate will be inserted in the probe forming aperture of the microscope to modify the electron beam before interaction with the sample as shown in Fig.~\ref{fig:intro}D. 

In order to correct aberrations, our target corrector function should contain linearly ramping phase profiles that closely match the aberration function of the probe, with a negative sign to cancel these aberrations out. The phase error tolerance of the design determines how far out of phase electrons can be after they are corrected. Here, the tolerance is set as $\pi/4$ following conventional aberration correction criteria \citep{kirkland-parameters, weyland2020tuning}.  Balancing spherical aberrations with defocus, we can minimize the curvature of the aberration function before applying the correction to improve the performance of the device.  Starting with no defocus, we show the fit aberration function (Fig.~\ref{fig:scherzer}D) and the residuals of a corrector (Fig.~\ref{fig:scherzer}E). We plot the corresponding curves in Fig.~\ref{fig:scherzer}D\&F for a corrector that includes defocus. Using a realistic crossbar width in the calculation, we can extend our maximum scattering angle from $q = 0.37$\AA$^{-1}$ to $q = 0.43$\AA$^{-1}$ by adding defocus. Note that both of these are larger maximum scattering angles than could be achieved with the ideal Scherzer condition alone (0.21\AA$^{-1}$).

%With a realistic cross bar width of $q_{gap} = 0.03 $\AA$^{-1}$, we can extend our maximum scattering angle from $q = 0.37$\AA$^{-1}$ to $q = 0.43$\AA$^{-1}$.

Fig.~\ref{fig:scherzer}G shows the same Scherzer probe as in Fig.~\ref{fig:scherzer}B for reference. We can compare this profile to the corrected probes. At lower spatial frequencies, the corrected probes both are less efficient than the Scherzer probe. However, the corrected probes reach a higher maximum spatial frequency, which means that they will improve the resolution of the microscope. The programmable phase plate that includes defocus achieves higher resolution, due to the overall more linear aberration surface. Similarly, Fig.~\ref{fig:scherzer}H shows the improved real space profiles. Although the central lobe is much smaller, the corrected probes have larger tails, as evident in the radii containing more than 95\% probe intensity.

The design of the aberration corrector plates resembles the device shown in Fig.~\ref{fig:blocked}A. An ideal corrector would be free of cross bars and support rings, but such a design is not physically realizable. The fill factor is defined as:
\begin{equation}
\textrm{fill factor} = \frac{\textrm{unblocked area}}{\textrm{blocked + unblocked area}}.
\label{eq:fill}
\end{equation}
A smaller fill factor will mean more of the electron beam is blocked. The cross bars were chosen to have 8-fold symmetry for more facile fabrication, but other support designs are possible. These parameters will ultimately depend on the physical size of the device and the resolution of the fabrication technique.

Fig.~\ref{fig:blocked}A shows the overall schematic of our design. Red and yellow lines indicate (+) and (--) terminals, and the black support bars are insulating.  Comparing Figs.~\ref{fig:blocked}A\&B highlights a key design choice in our programmable phase plate. In Fig.~\ref{fig:blocked}A we block the center of the phase plate, which means that every aperture can have two terminals. By comparison, the design in Fig.~\ref{fig:blocked}B has an open aperture in the center with a single electrode, as two electrodes in this circular aperture could not create a linearly ramping profile. The blocked center is the preferable choice, as it allows for more control over the electron beam profile. The color legend for this figure and the other probe plots throughout the paper is shown in Fig.~\ref{fig:blocked}C, and it shows that designs in both Figs.~\ref{fig:blocked}A\&B can largely remove aberrations in the probe. Figs.~\ref{fig:blocked}D\&E show the corresponding CTF and radial probe profiles for these two devices, illustrating the same degree of correction. 

Next, we consider two approaches for determining the width of each aperture for the corrector: (1) evenly spaced apertures and (2) adaptively spaced apertures. Evenly spaced apertures were designed such that all the aperture widths are the same. The size of the device is increased until the corrected phase profile in any aperture reaches the tolerance of $\pi/4$. The aberration function and residuals for this method are plotted in Figs.~\ref{fig:adaptive}A\&B respectively. For adaptively spaced apertures, the width of each ring is determined independently, and its size is maximized such that the corrected phase profile in any aperture is within the tolerance of $\pi/4$ (Figs.~\ref{fig:adaptive}A\&C). Comparing the residuals in Figs.~\ref{fig:adaptive}B\&C we see why the adaptive corrector is the better choice. For the same number of rings, we achieve a higher maximum scattering angle.

Fig.~\ref{fig:adaptive}A highlights that the optimal defocus is not the same for these two designs, leading to different aberration function profiles. The challenge of finding the optimal defocus is difficult to solve analytically. This is especially true in the adaptive corrector cases, as the profile of maximum achievable scattering angle for various focus values oscillates. The optimal defocus for these devices was determined by starting at $C_1 = 0$ and progressively updating the defocus and calculating the maximum scattering angle. Once we reach a local maximum in scattering angle, we defined the best defocus and aperture size based on this maximum (Fig.~\ref{fig:adaptive}D). Fig.~\ref{fig:adaptive}E shows the maximum achievable scattering angle for both approaches for different numbers of rings.  With one ring, these two approaches yield the same degree of correction. However, for more rings, the adaptive corrector consistently performs better. 

We next evaluate the parasitic contributions of probe tails for the two approaches. Fig.~\ref{fig:adaptive}F shows the radii containing 50\% and 95\% of the cumulated probe intensity. The 50\% radius most closely corresponds to degree of correction or attainable resolution. As we add more rings to the corrector, we reach a higher scattering angle of corrected electrons, so the real space probe size improves. The adaptive corrector shows improved resolution over the evenly spaced corrector. The 95\% intensity shows another challenge with the evenly spaced corrector, namely broader probe tails. For a 4-ring device, we can plot these same metrics against fill factor using Eq.~\ref{eq:fill}.  As the fill factor gets lower the resolution stays nearly constant (50\% intensity), while the probe tails, as represented by 95\% intensity, increase.

To understand the limitations of the programmable corrector, we can also look at the profiles of the probes in real space. Fig.~\ref{fig:probe_profile} shows the profile of the probe along electron beam propagation axis for (1) a Scherzer condition probe according to Eq.~\ref{eq:c1}, (2) an evenly spaced corrected and (3) adaptively spaced corrected probe in A-C respectively. The 50\% and 95\% radial profiles of these probes are plotted in Fig.~\ref{fig:probe_profile}D\&E as well. While the corrected probes are much narrower, indicating a state of aberration correction, they have modes above and below the sample plane, which are similar to x-ray beam profiles in Fresnel zone plate experiments~\citep{attwood2000soft}.  These extra modes would probe specimen information above and below the sample, which is an especially important consideration for thicker samples.

\begin{figure}
  \begin{center}
    \includegraphics[width=0.45\textwidth]{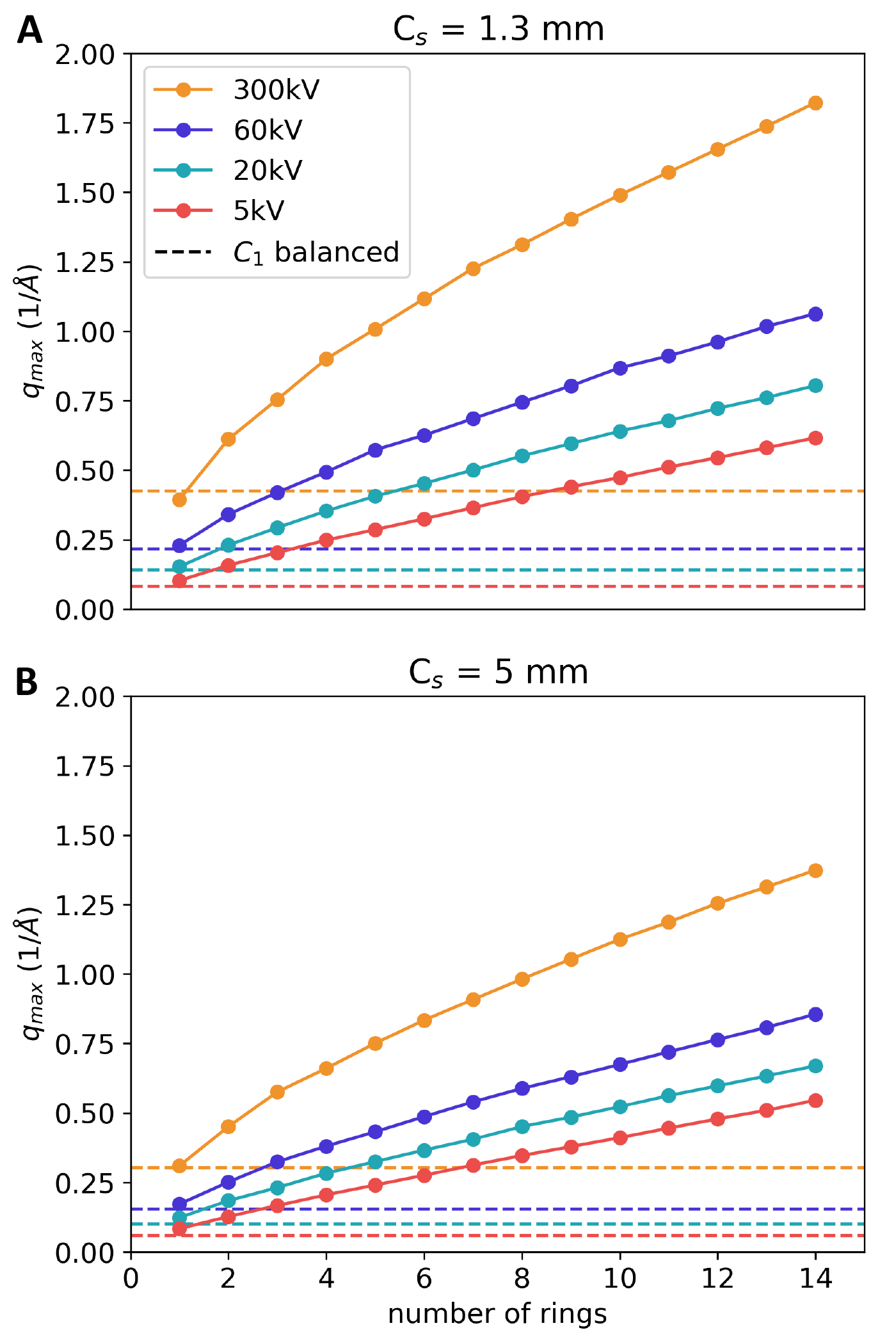}
  \end{center}
  \caption{Simulations of maximum scattering angle as a function of the number of rings for different voltages with (A) 1.3mm $C_3$ and (B) 5mm $C_3$ aberrated probes.}
  \label{fig:sem}
\end{figure}

Fig.~\ref{fig:sem} explores the performance of the device at various accelerating voltages at two values for spherical aberration: 1.3mm, which is the range of spherical aberration coefficients for a high voltage S/TEM~\citep{hong2021multibeam} and 5mm, which is closer to the range for a low voltage Scanning Electron Microscope (SEM)~\citep{zach1995aberration, joy2008aberration}.  The horizontal dashed lines represent the $q_{max}$ achievable with balancing $C_1$ (Eq.~\ref{eq:c1}), while the solid profiles show the maximum achievable $q_{max}$ for an adaptively spaced corrector of various number of rings. This figure underscores the choice of 60kV as a goal for achieving atomic resolution. At 300 kV, the $q_{max}$ is approaching 0.5A$^{-1}$, which would provide excellent atomic resolution imaging.

We also consider the possibility of using this type of corrector for an SEM. The results in Fig.~\ref{fig:sem} suggest that although it may be possible to approach atomic resolution at 20kV and 5kV with a programmable phase plate, it would require many more apertures than at 60kV, making the implementation of a programmable phase plate for aberration correction in an SEM challenging. Moreover, despite the spherical aberrations in an SEM being higher than in a STEM, the chromatic aberrations are more problematic. Thus there may be other challenges to achieving atomic resolution which this electrostatic phase plate corrector cannot address~\citep{zach1995aberration, joy2008aberration}. 

\section*{Methods}

Unless otherwise noted, theory simulations are at 60kV accelerating voltage with a $C_3$ of 1.3mm.  Low voltage S/TEM (60kV) is intrinsically lower in resolution than higher voltage experiments, creating more incentive for aberration correction. This spherical aberration coefficient was chosen based on realistic microscope parameters~\citep{hong2021multibeam}.

The graphene and silicon structures for these simulations were built from files available through the Materials Project~\citep{Jain2013}. STEM simulations were performed using the \textit{ab}TEM~\citep{madsen2021abtem} multislice code based on methods laid out by ~\cite{kirkland-parameters}. Because two-dimensional materials need a high number of frozen phonon (FP) configurations to converge~\citep{dacosta2021prismatic}, twisted graphene bilayer simulations were run with 50 FPs, while tetracutinase and Si simulations were run with 12 FPs, and the standard deviation of the displacement was 0.1 \AA. For the Moir\'e and Si simulations, the dark field detector was integrated from 80-135 mrad. Dark field (DF) and differential phase contrast (DPC) reconstructions were performed in \textit{ab}TEM.

Simulations of tetracutinase~\citep{parker2022scanning} were performed in \textit{ab}TEM. The probe was defocused to about 10 nm, and 4D datasets were simulated with a 1 nm step size to ensure sufficient probe overlap. Poisson noise was added to simulate a dose of 500 e$^-$/\AA$^{2}$. Ptychography refers to a family of phase retrieval techniques, where the phase of the sample and the probe are reconstructed from 4D-STEM datasets. Ptychographic reconstructions used the regularized ptychographic iterative engine algorithm as implemented in \textit{ab}TEM \citep{maiden2017further}. 15 iterations were run with a step size of 0.1. 

The Fourier Ring Correlation (FRC) is used to compute spatial frequencies transferred in image reconstructions~\citep{van2005fourier, banterle2013fourier}.
\begin{equation}
    \textrm{FRC}(r) = \frac{\sum_{r_i \in r}  F_1(r) \cdot F_2(r)^*}{\sqrt{\sum_{r_i \in r}  F_1^2(r) \cdot \sum_{r_i \in r}  F_2^2(r) }},
\end{equation} 
where $F_1$ and $F_2$ are the Fourier transforms of two real space reconstructions and $*$ denotes the complex conjugate.  Four reconstructions were calculated for each condition, and the FRCs were averaged to improve signal to noise. The half bit criteria define the signal to noise ratio at a given frequency, and the first intersection of the half bit curve (F(r)) with the FRC(r) with a negative slope of the difference curve defines the resolution~\citep{van2005fourier, banterle2013fourier}. The half-bit criteria are defined as

\begin{equation}
    \textrm{F}(r) = \frac{2}{\sqrt{\textrm{N}(r)/2}},
\end{equation}

where N(r) is the number of pixels in each ring. 

\section*{Results and Discussion}

\begin{figure*}
  \begin{center}
    \includegraphics[width=\textwidth]{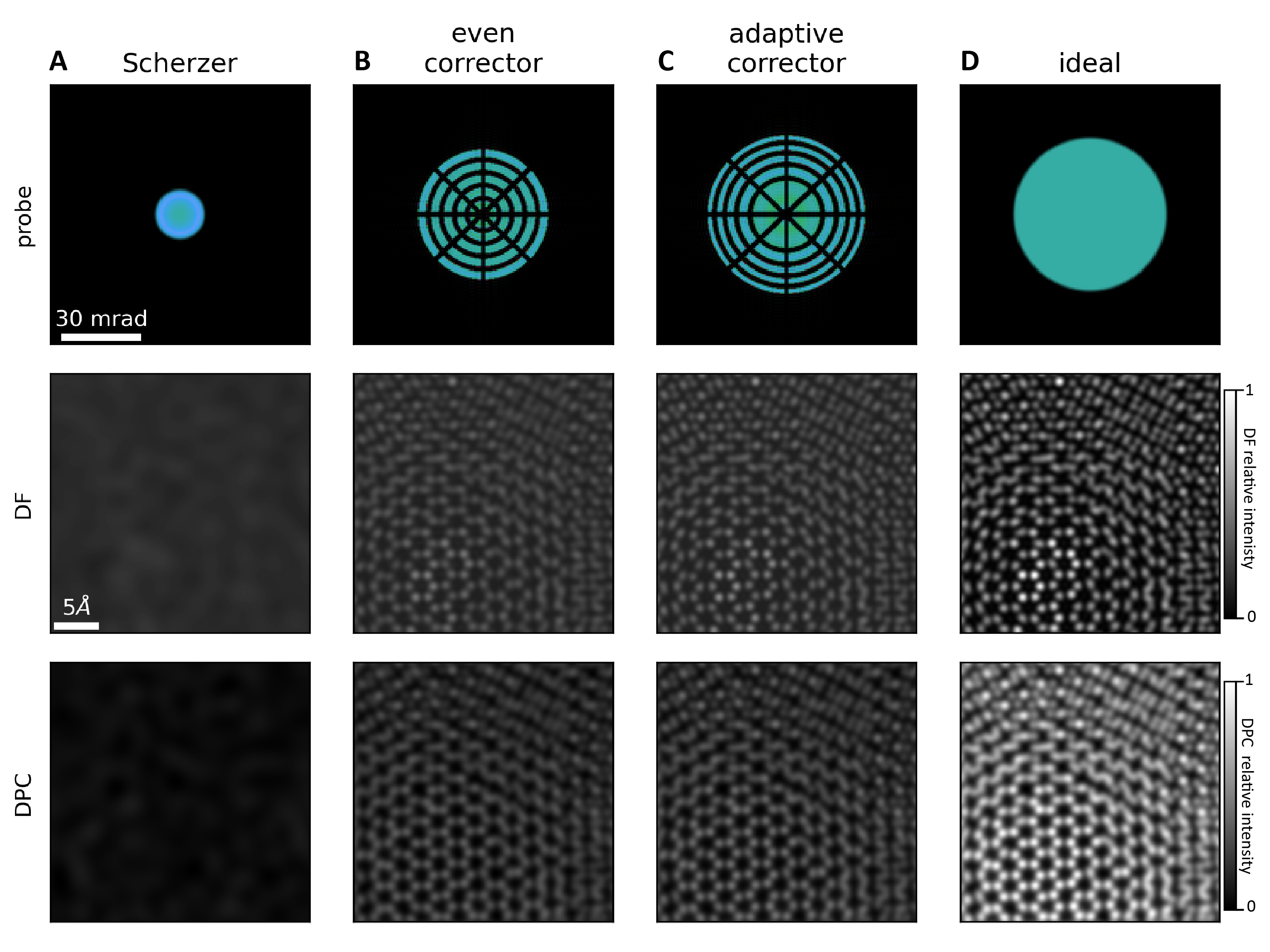}
  \end{center}
  \caption{(A) The Scherzer condition probe does not provide sufficient resolution for imaging of twisted bilayer graphene, while the (B) evenly spaced and (C) adaptively spaced corrected probes provide atomic resolution images. Compared to an (D) ideal probe the efficiency is reduced in the corrected images as shown by the lower contrast. 60kV simulations with $q_{max}=0.22$\AA$^{-1}$, $q_{max}=0.51 $\AA$^{-1}$, $q_{max}=0.63$\AA$^{-1}$ and $q_{max}=0.63$\AA$^{-1}$ in A-D respectively.}
  \label{fig:moire}
\end{figure*}

\begin{figure}
  \begin{center}
    \includegraphics[width=0.45\textwidth]{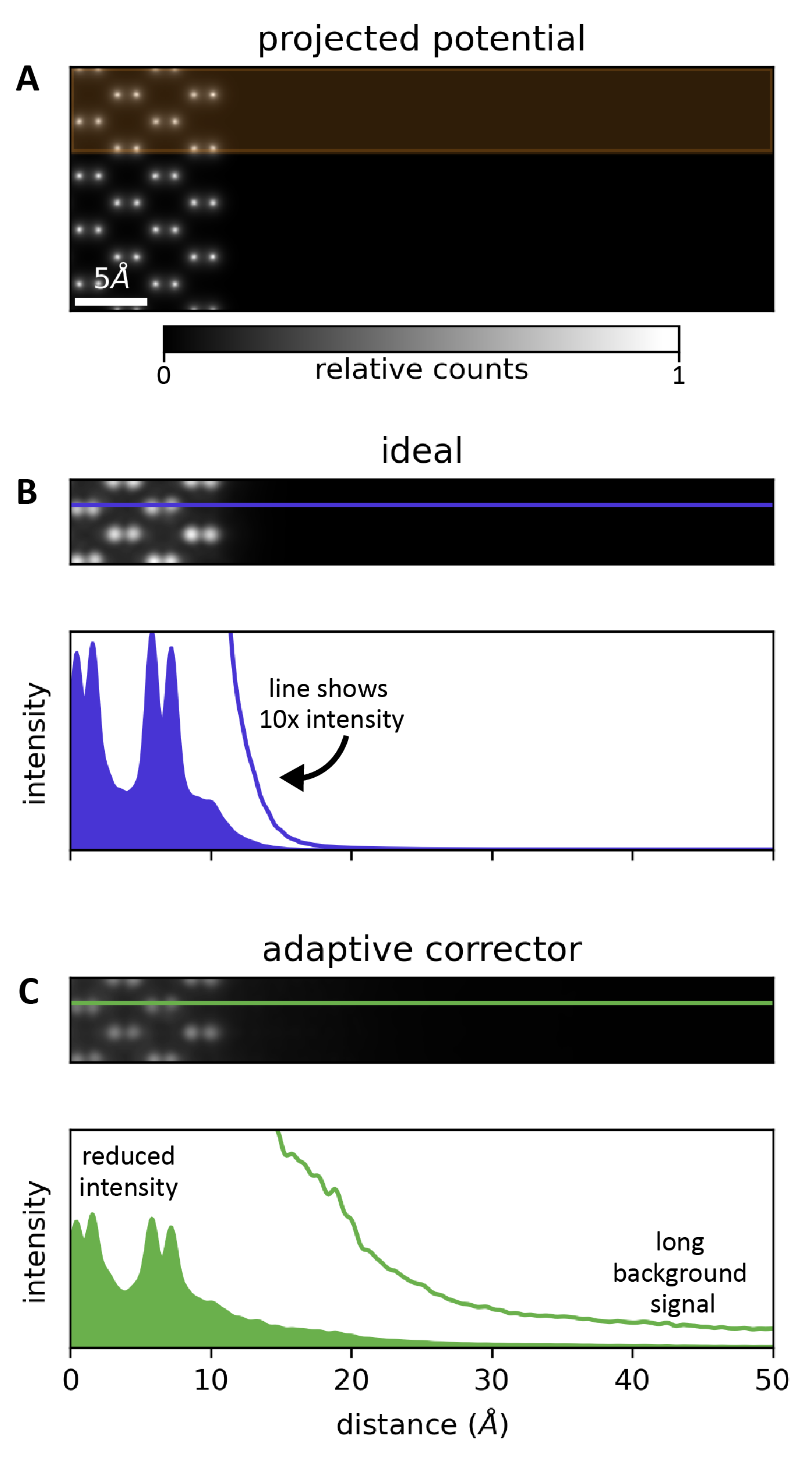}
  \end{center}
  \caption{(A) Projected potential of a 20 nm Si sample. Comparing the (B) perfect probe and (C) corrected probe, we observe that despite the benefits of the programmable phase plate for improving resolution, there is a cost of reduced intensity and strong background signal.}
  \label{fig:silicon}
\end{figure}

\begin{figure}
  \begin{center}
    \includegraphics[width=0.45\textwidth]{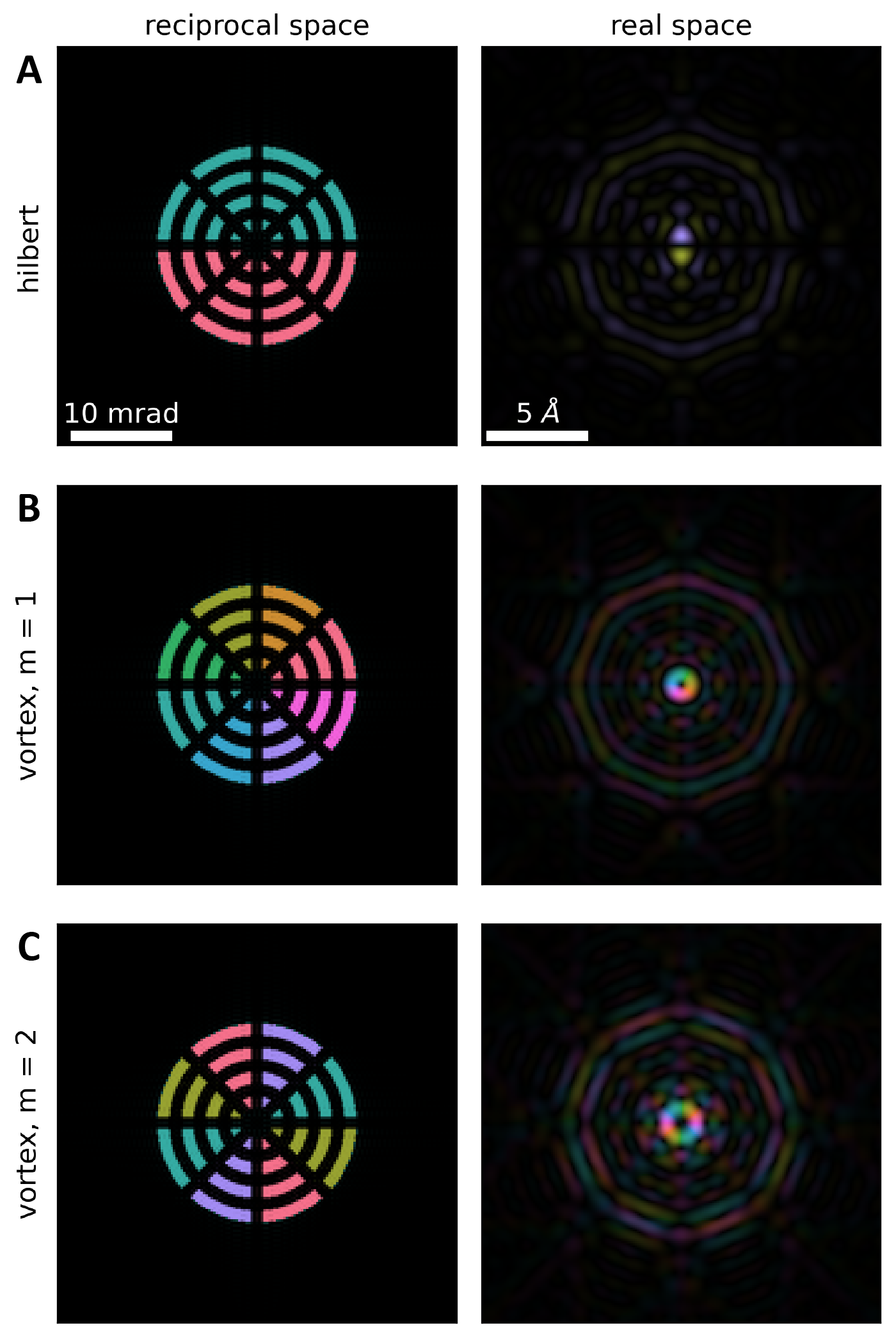}
  \end{center}
  \caption{A programmable phase plate allows for the implementation of more exotic beam profiles including (A) a probe with a Hilbert plate or a vortex beam with (B) $m=1$ or (C) $m=2$}
  \label{fig:vortex}
\end{figure} 

\begin{figure*}
  \begin{center}
    \includegraphics[width=\textwidth]{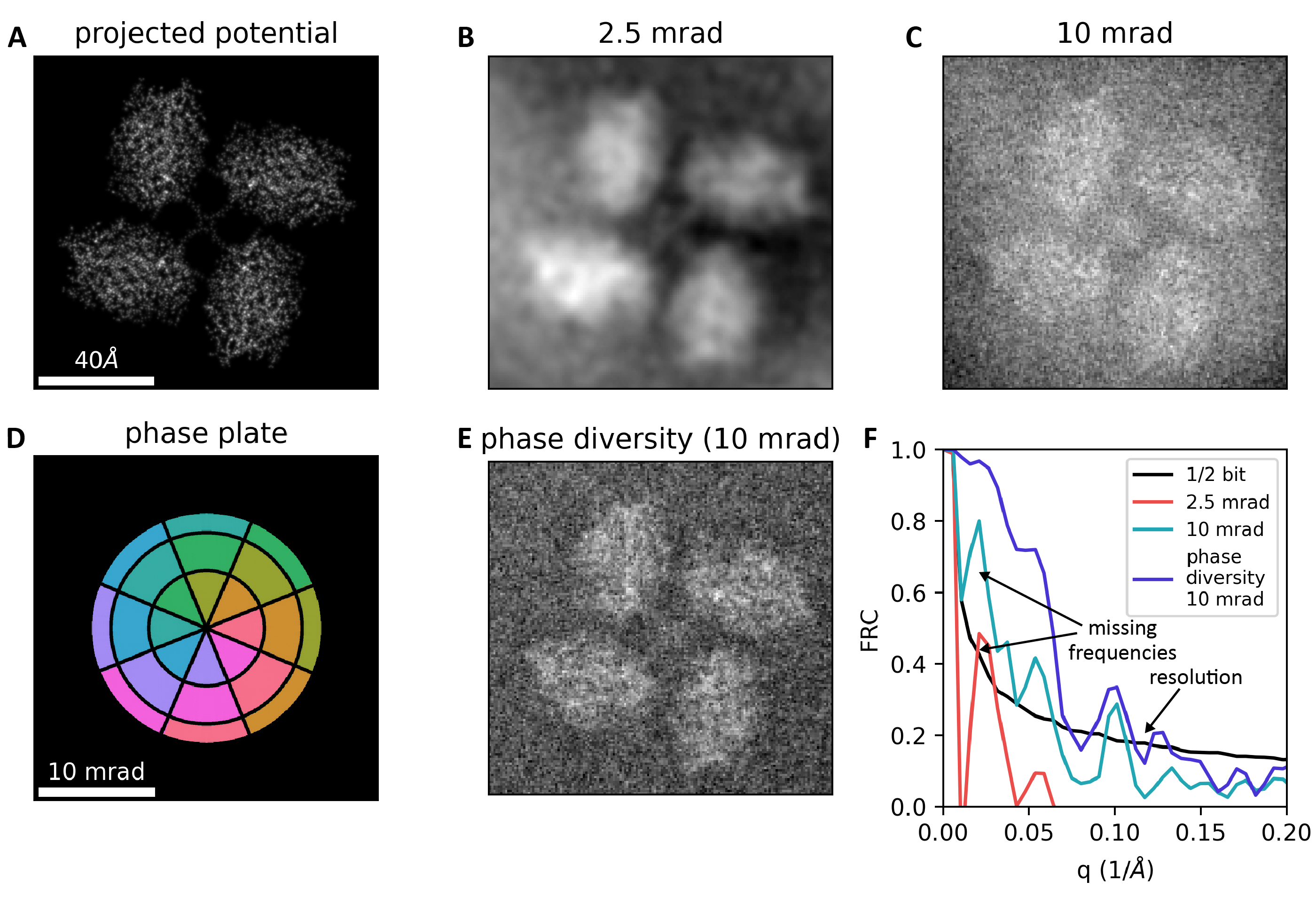}
  \end{center}
  \caption{(A) Projected potential of tetracutinase and defocused ptychography reconstructions with (B) 2.5 and (C) 10 mrad defocused probes. A (D) phase plate adds diversity to the (E) reconstructions, which capture a wider range of spatial frequencies. (F) FRC profiles evaluate transfer of information in these experiments.}
  \label{fig:ptycho}
\end{figure*} 

We use a Moir\'e graphene bilayer structure as a model system to test the performance of programmable phase plate as an aberration corrector. Fig.~\ref{fig:moire}A shows a dark field and Differential Phase Contrast (DPC) image of the Moir\'e lattice under optimal conditions as described by Eq.~\ref{eq:c1}. Because of the limited resolution of the probe in reciprocal space, it is not possible to resolve the fine features in the twisted graphene structure. However, Fig.~\ref{fig:moire}B\&C shows simulations of the same structure with the evenly and adaptively spaced correctors. The adaptively spaced corrector performs slightly better than its evenly spaced counterpart, as it can reach higher maximum scattering angles($q_{Scherzer}=0.22$\AA$^{-1}$, $q_{even}=0.51 $\AA$^{-1}$, $q_{adaptive}=0.63$\AA$^{-1}$). 

We compare these results to an ideal probe with  $q_{max}=0.63$\AA$^{-1}$ (Fig.~\ref{fig:moire}D). Although the resolution is similar to the images in Fig.~\ref{fig:moire}B\&C, there is stronger contrast in the lattice.  The CTFs in Fig.~\ref{fig:scherzer} show reduced transfer of information with a phase plate corrector as compared to an ideal probe, especially at low spatial frequencies, so these images are in good agreement with the CTF calculations. 

The limitations of the programmable phase plate approach to aberration correction are more clearly shown with a thicker sample. In Fig.~\ref{fig:silicon}, we explore the impact of aberration correction on a 20nm thick silicon sample oriented along the $[11\overline{1}]$ zone axis. Fig.~\ref{fig:silicon}A shows the projected electron potential of the sample with the area of interest highlighted. We include a large vacuum region next to the sample, to highlight the impact of the long probe tails. In Fig.~\ref{fig:silicon}B, we show the image and line profile from a simulation with a perfect probe of scattering angle $q_{max} = 0.63 $\AA$^{-1}$. We can clearly resolve the dumbbells and the intensity drops off within a few \AA{}ngstr\"oms of the edge of the sample. 

We can compare these results to an image simulation with spherical aberrations and the corrector. We use a 5-ring programmable phase plate with the same $q_{max}$ as Fig.~\ref{fig:silicon}B. As shown previously, we would not be able to achieve atomic resolution for this aberrated probe without correction. With this corrector, we can successfully resolve the dumbbells in this sample. However, due to the long probe tails we see a strong background signal extending well into the vacuum. Even within 4 nm of the sample, the intensity has not dropped to zero, shown by the line profiles.  Although the nominal central lobe size is sub-\AA{}ngstr\"om, the probe is sensitive to information nanometers away from the main area of illumination.  Moreover, there is significantly reduced intensity at the atomic sites, similar to Fig.~\ref{fig:moire}. These probe tail features lower the contrast more in thicker samples due to beam broadening during beam-specimen interactions. 

Here we have described how the programmable phase plate can be used for spherical aberration correction. This hardware also allows for high flexibility in defining the electron beam wave function. For example, a Hilbert plate, where half of the beam is phase shifted by $\pi$ can be used in TEM to improve contrast in studies of weak phase objects~\citep{danev2002novel}. Such a beam profile can be implemented with the programmable phase plate as shown in Fig.~\ref{fig:vortex}A.  There are many other types of phase plates that are more routinely used in TEM~\citep{malac2021phase}, but have been little explored in STEM. A programmable phase plate would allow for more testing of these usual beam profiles.

Vortex beams in S/TEM has been proposed for probing a number of material properties including chirality ~\citep{juchtmans2015using, harvey2015demonstration, beche2017efficient}, symmetry ~\citep{juchtmans2016extension,ribet2022defect}, and magnetic structure \citep{verbeeck2010production, rusz2013boundaries, grillo2017observation}. Fig.~\ref{fig:vortex} shows how these beams can be implemented in real and reciprocal space.  The wave function of a vortex beam, $\Psi_v(q)$, is defined by:
\begin{equation}
\Psi_v(q) = \Psi(q)e^{im\phi} 
\label{eq:vortex}
\end{equation}
Here,$\Psi(q)$ is the wave function of an unmodified beam, m is the quantum number, and $\phi$ is the azimuthal coordinates with respect to the propagation direction of the electron beam. Energy-loss magnetic circular dichroism (EMCD) experiments, for example, probe magnetic order with a chiral vortex beam ~\citep{schattschneider2006detection, verbeeck2010production}. In such an experiment, one needs to compare an electron energy loss experiment from the same area with a $m=+1$ and $m=-1$ beam, which can be challenging to implement experimentally. A programmable phase plate can be used to make a vortex beam, as shown in Fig.~\ref{fig:vortex}B, and the tunable nature of this design lends itself to EMCD experiments. 

Finally, our programmable phase plate could be used for adding phase diversity in a STEM probe for ptychography experiments. Ptychography experiments are often performed in a defocused probe configuration, which adds phase in the incident beam~\citep{rodenburg2019ptychography}. However, there have been a variety of studies that have suggested that adding additional phase diversity into the probe, including dynamic phase, can help with more efficient reconstructions, especially at low spatial frequencies~\citep{candes2015phase, yang2016enhanced, pelz2017low, allars2021efficient}.  A programmable phase plate would be well suited to add phase into the probe for ptychography experiments.

%\cite{ophus2016efficient} showed how incorporating a phase plate with zone of 0 and $\pi/2$ phase in a STEM leads to improved transfer of information for heavy and light elements across a wide-range of spatial frequencies in 4D-STEM experiments. \cite{yang2016enhanced} utilized this phase plate for ptychography reconstructions, showing how ptychographic reconstructions were improved by adding phase diversity. In these experiments, the diversity in the probe came from the phase plate, so the ptychography data sets were collected in focus. Therefore, the reconstructed dark field images from the same data set did not suffer from defocus. 

%In ptychography experiments using defocused probes to probe biological molecules,~\cite{pelz2017low} demonstrated that a probe with random phase shifts can improve the signal to noise in reconstructions. Similarly, \cite{allars2021efficient} showed how ptychography reconstructions in a near-field TEM set-up with structured illumination can efficiently capture the phase from a large area with a few number of diffraction patterns.  In an extreme ultraviolet ptychography experiment, vortex beams can improve ptychographic reconstructions for periodic structures~\citep{wang2023high}.  \cite{candes2015phase} have also demonstrated how adding structured illumination to a beam before each diffraction pattern is acquired can create a coded data set to help a ptychographic reconstruction converge to a global minimum. A programmable phase plate could allow for dynamic electron beam modification during a scan.

Using tetracutinase a model system, we explore this phase plate configuration, and in particular how adding phase diversity can be used to improve the information transfer for defocused probe ptychographic reconstructions. Similar to many biological structures, these megamolecules are weak phase objects and beam sensitive, making low dose ptychography a good approach to imaging these materials~\citep{parker2022scanning}. Fig.~\ref{fig:ptycho}A shows the projected potential of this small four-lobed molecule. The reconstructions using an ideal probe with 2.5 mrad and 10 mrad convergence angles (Fig.~\ref{fig:ptycho}B-C respectively) show the impact of convergence angle on the reconstruction. At a smaller convergence angle, the low spatial frequency information is captured, while the larger convergence angle results in an image that appears high-pass filtered. The FRC plots (Fig.~\ref{fig:ptycho}E) show improved resolution for the larger convergence angle. These results are in good agreement with the images and experimental results from the literature~\citep{zhou2020low}. 

Fig.~\ref{fig:ptycho}D shows a STEM phase plate, which we use for this simulation. This design was chosen to incorporate phase diversity in the probe. The ptychographic reconstruction using this probe wave function is shown in Fig.~\ref{fig:ptycho}E. The image shows a wider range of spatial frequencies -- the low spatial frequency information is preserved while capturing the fine detail in the megamolecule's lobes. In addition, these reconstructions are more robust in the vacuum region to systematic errors. The FRC profile for this reconstruction shows the same resolution as compared to the conventional 10 mrad probe, but with improved transfer at lower spatial frequencies. Overall, these simulations suggest that incorporating phase diversity into an incident probe can help improve ptychographic reconstructions.

\section*{Conclusion}

In this work we have shown how to design a programmable phase plate for spherical aberration correction in light of realistic design criteria. We have illustrated how this device can be used to correct third-order spherical aberrations to produce atomic-resolution images, and a similar approach could be used to correct higher-order spherical aberrations. One of the key limitations of these devices is the long probe tails that arise due to supporting cross bars in the device. The probe tails are especially problematic in thicker specimens, resulting in signals when the probe is outside of the specimen. A programmable phase plate can also be used to create more complex probe profiles, such as a vortex beam.  We have shown how these types of probe profiles can add phase diversity into the probe to improve transfer of information in ptychographic reconstructions. 

\section*{Competing interests}
The authors declare that they have no competing interests.

\section*{Acknowledgements} \label{sec:acknowledgements}
This material is based upon work supported by the U.S. Department of Energy, Office of Science, Office of Workforce Development for Teachers and Scientists, Office of Science Graduate Student Research (SCGSR) program. The SCGSR program is administered by the Oak Ridge Institute for Science and Education for the DOE under contract number DE‐SC0014664. SMR acknowledges support from the IIN Ryan Fellowship and the 3M Northwestern Graduate Research Fellowship.  SEZ was supported by the National Science Foundation under STROBE Grant No. DMR 1548924. CO acknowledges support from the US Department of Energy Early Career Research Program. This material is based upon work supported by the National Science Foundation under Grant No. DMR-1929356.  This work made use of the EPIC facility of Northwestern University’s NU\textit{ANCE} Center, which has received support from the SHyNE Resource (NSF ECCS-2025633), the International Institute of Nanotechnology (IIN), and Northwestern's MRSEC program (NSF DMR-1720139).  Work at the Molecular Foundry was supported by the Office of Science, Office of Basic Energy Sciences, of the U.S. Department of Energy under contract number DE-AC02-05CH11231. 

\section*{References} 
% \section*{Supplemental}
% \setcounter{figure}{0}
% \makeatletter 
% \renewcommand{\thefigure}{S\@arabic\c@figure}
% \makeatother

% \bibliographystyle{MandM}
\bibliography{refs}    

\end{document}